\documentclass[%
 reprint,
 amsmath,amssymb,
 aps,prl,
]{revtex4-2}

\usepackage{graphicx}
\usepackage{dcolumn}
\usepackage{bm}
\usepackage{xcolor}

\begin{document}
\preprint{APS/123-QED}
\title{Multimodal Anti-Reflective Coatings\\
 for Perfecting Anomalous Reflection from Arbitrary Periodic Structures}
\author{Sherman W. Marcus}
\email{shermanm@technion.ac.il}
\author{Vinay K. Killamsetty}
\author{Ariel Epstein}

\affiliation{Andrew and Erna Viterbi Faculty of Electrical and Computer Engineering\\ Technion - Israel Institute of Technology, Haifa 3200003, Israel}

\date{\today}


\begin{abstract}
Metasurfaces possess vast wave-manipulation capabilities, including reflection and refraction of a plane wave into non-standard directions. This requires meticulously-designed sub-wavelength meta-atoms in each period of the metasurface which guarantee unitary coupling to the desired Floquet-Bloch mode or, equivalently, suppression of the coupling to other modes. Herein, we propose an entirely different scheme to achieve such suppression, alleviating the need to devise and realize such dense scrupulously-engineered polarizable particles. Extending the concept of anti-reflective coatings to enable simultaneous manipulation of multiple modes, we show theoretically and experimentally that a simple superstrate consisting of only several uniform dielectric layers can be modularly applied to \textit{aribtrary} periodic structures to yield perfect anomalous reflection. This multimodal anti-reflective coating (MARC), designed based on an analytical model, presents a conceptually and practically simpler paradigm for wave-control across a wide range of physical branches, from electromagnetics and acoustics to seismics and beyond.   
\end{abstract}


\maketitle


Metasurfaces have been shown to possess wave-control capabilities not previously available, both for transmission through them and for reflection from them \cite{HollowayIEEEAPMagPhysRep2012,TretyakovPhilTrans2015,GlybovskiPhysRep2016}. For many applications, the periodic metasurface is divided into discrete sub-wavelength cells (meta-atoms) which are meticulously designed to provide the desired scattered fields. A higher density of meta-atoms generally provides fields closer to the design goals at the expense of increased design and fabrication complexity. The wave-control capabilities are often demonstrated by designs which reflect \cite{SunNatMat2012,AsadchyPRL2015} or transmit \cite{LalanneOptLett1998,LalanneOptSocAmA1999,HasmanAPL2003,YuScience2011,PfeifferPRL2013,MonticonePRL2013,SelvanayagamOptexp2013} an incident plane wave into a single non-standard direction (anomalous reflection or refraction), or into multiple non-standard directions (beam-splitting)  \cite{EpsteinPRL2016}.  

Huygens’ metasurfaces (HMSs) \cite{PfeifferPRL2013,MonticonePRL2013,SelvanayagamOptexp2013} successfully produced anomalous refraction for moderate refraction angles, but at extreme angles wave-impedance mismatch led to the additional presence of a specularly reflected wave \cite{SelvanayagamOptexp2013,EpsteinIEEEAP2014}.  This failure was overcome by omega bi-anisotropic metasurfaces (OBMSs) which introduced an additional magnetoelectric degree of freedom, leading to a metasurface capable of producing \textit{perfect} anomalous reflection, refraction and beam-splitting \cite{EpsteinPRL2016,WongIEEEAWPL2015,AsadchyPRB2016,EpsteinIEEETAP2016}; perfect in the sense that all incident power is transferred to the desired anomalously scattered wave(s).  Comparing analytical solutions for both the HMS \cite{MarcusPhysRevB2019} and the OBMS \cite{MarcusPhysRevB2020}, we recently employed a ray-optics interpretation of the process to demonstrate the equivalence between an OBMS, and a HMS with a virtual anti-reflective coating (ARC) \cite{MarcusPhysRevB2020,MarcusMM2020}.  The purpose of the ARC (be it virtual \cite{MarcusPhysRevB2020} or real \cite{RautEnvSci2011}) is, of course, to suppress a single, unwanted specularly reflected mode, which is done by ensuring the resulting multiply-reflected waves add up destructively. 

\begin{figure}[t]
  \begin{center}
  \includegraphics[width=3.5in]{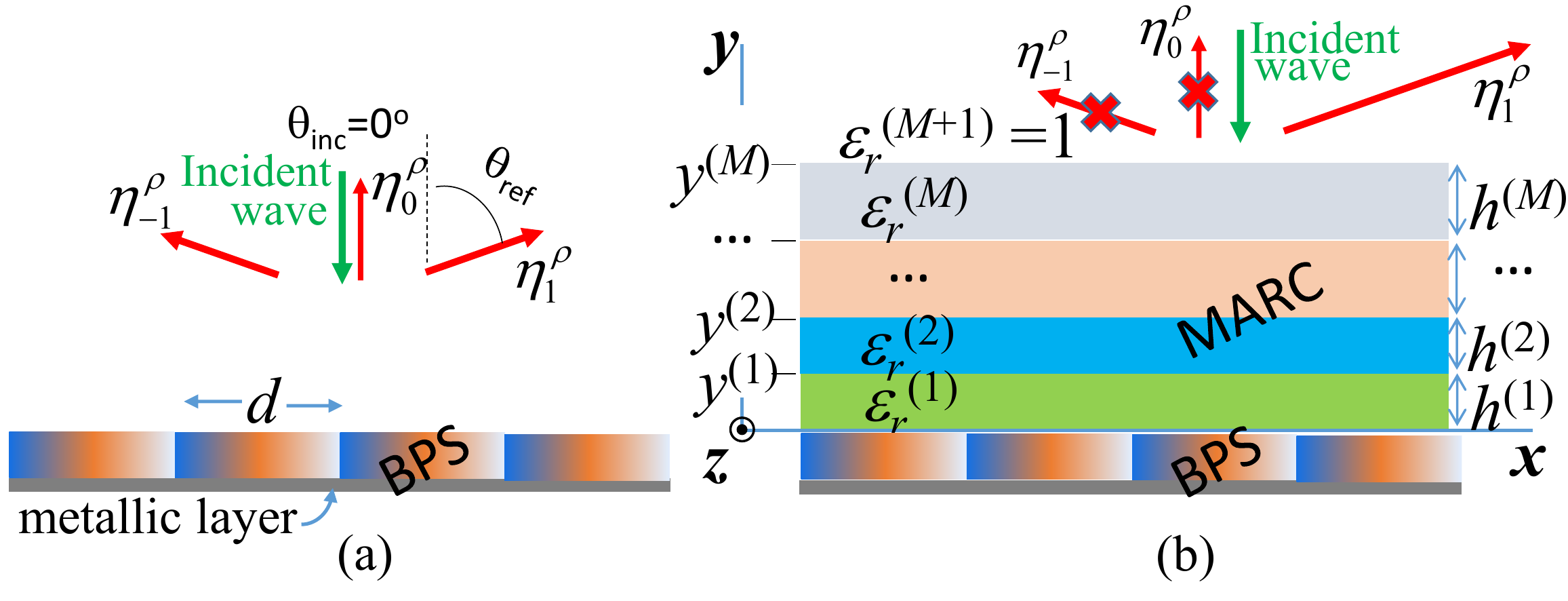}
  \caption{A $d$-periodic, impenetrable, simple basic periodic surface (BPS) for reflecting a normally incident wave into directions $\pm \theta_\text{ref}$ and $0^{\circ}$.  (a) Without the MARC, the amplitudes of these FB reflected waves are finite. (b) With a properly designed MARC, all but the desired FB wave are suppressed.  The MARC consists of $M$ uniform dielectric layers; the thickness of layer $m$ is $h^{(m)}$, its upper surface is at $y=y^{(m)}$, and its relative permittivity is $\epsilon_r^{(m)}$. $\eta^{\rho}_p$ is the relative power coupled to reflected mode $p$.} \label{FigIntro2}
  \end{center}
\end{figure}

In this Letter, we suggest and demonstrate experimentally that the ARC principle of suppressing a single mode can be broadened to suppressing \textit{multiple} modes, which could open up a myriad of novel wave-control possibilities (Fig. \ref{FigIntro2}).  This is tantamount to enhancing the other remaining modes of interest, a task previously performed by the discrete sub-wavelength meta-atoms of the metasurface.  This new \textit{multimodal} anti-reflective coating (MARC), which consists of only several \textit{uniform} layers of dielectrics (Fig. \ref{FigIntro2}), would replace many \textit{periodically distributed} meta-atoms, thereby significantly simplifying the structure, design and implementation. Extending the underlying physical mechanism of the standard ARCs, the dielectric stack should be engineered to ensure destructive interference of all unwanted modes, while fully accounting for the mode-coupling characteristics of the periodic surface to which the MARC is attached. Such a device offers many advantages over the aforementioned state-of-the-art metasurfaces. First, it can be applied to \textit{any arbitrary} surface with the required period (termed herein "basic periodic surface" - BPS), including surfaces that have not undergone specialized design procedures; in fact, they do not need to adhere to the homogenization approximation conditions \cite{Kuester2003,Tretyakov2003}, thus can be much sparser than usual. Second, it can be applied \textit{after-the-fact} to perfect the performance of suboptimal structures, being an independent superstrate module employed without any modification to the original BPS. Lastly, and perhaps most importantly, the elementary wave physics employed in propagation through layered media, on which MARC synthesis relies, makes this solution applicable to a wide variety of physical systems, ranging from electromagnetic and optical devices to acoustic components and seismic scenarios.

To facilitate this concept and showcase its merits, we begin by devising an efficient analytical model for determining the characteristics of the uniform dielectric layers of the MARC needed to obtain desired anomalous effects. This model would calculate the fields scattered from the BPS-MARC combination for different layer configurations, and would employ parametric variation schemes to choose an optimum configuration. Consider, then, a BPS that is invariant in the $z$-direction with periodicity $d$ along $x$, coated below by a metallic layer to assure reflection (Fig. \ref{FigIntro2}). For transverse magnetic (TM) polarization, the magnetic field $\textbf{H}=H_z \hat{\textbf{z}}$, while the electric field \textbf{E} will not contain a $z$-component. The BPS is coated above by a MARC consisting of $M$ dielectric layers, where layer $m$ is characterized by dielectric constant $\epsilon_r^{(m)}$ and thickness $h^{(m)}=y^{(m)}-y^{(m-1)}$.  In accordance with the Floquet-Bloch (FB) theorem, a plane wave $H_{z,\text{inc}}(x,y)=H_0e^{ik_{x0}x}e^{-ik_{y0}^{(M+1)}y}$ incident in region $M+1$  on the BPS-MARC system at an angle $\theta_{\text{inc}}$ will be scattered into a discrete spectrum of (propagating and evanescent) waves: 

\begin{equation}\label{EqFB1}
   H_z^{(M+1)}(x,y)=H_{z,\text{inc}}+\sum\limits_{p=-\infty }^{\infty } \rho_{p}{e^{ik_{xp}x}}{{e}^{ik^{(M+1)}_{yp}y}},
\end{equation}
where $y>y^{(M)}$, and the transverse and longitudinal wavenumbers of the $p$th mode in the $m$th layer are given, respectively, by ${k}_{xp}=k\sin {{\theta }_\text{inc}}+2p\pi/d$, $k^{(m)}_{yp}=(k^2\epsilon_r^{(m)}-k_{xp}^2)^{1/2}$.  $d=\lambda/|\sin\theta_\text{ref}-\sin\theta_\text{inc}|$ is the BPS periodicity, $\theta_\text{ref}$ is the design  anomalous reflection angle, the wave number $k=2\pi/\lambda$, $\lambda$ is the wavelength in free space, an $e^{-i\omega t}$ time dependence is assumed, and the modal reflection coefficients $\rho_p$ are, as yet, unknown. 

In order to keep the geometry simple, normal incidence $(\theta_\text{inc}=0)$ will be assumed, along with $\theta_\text{ref}>30^\circ$ which leads to three values of $p$ for which $k_{yp}$ is real: $p=[-1,0,1]$ \cite{EpsteinPRL2016}, corresponding to waves propagating towards $\theta_p=\arctan(k_{xp}/k_{yp}^{(M+1)})=[-\theta_{\text{ref}},0,\theta_{\text{ref}}]$ (red arrows in Fig. \ref{FigIntro2}); all other terms in Eq. \eqref{EqFB1} represent evanescent waves. The goal of the analytical model is to determine the efficiencies $\eta^\rho_p=|\rho_p|^2\cos\theta_p/\cos\theta_{\text{inc}}$ with which power is coupled from the incident wave to each of the scattered waves. Straightforward optimization algorithms can then be employed to determine, say, the $h^{(m)}$ for which perfect anomalous reflection is attained: $\eta^\rho_{-1} \rightarrow 0$,  $\eta^\rho_{0} \rightarrow 0$, $\eta^\rho_{1} \rightarrow 1$.

\begin{figure}
  \begin{center}
  \includegraphics[width=3.5in]{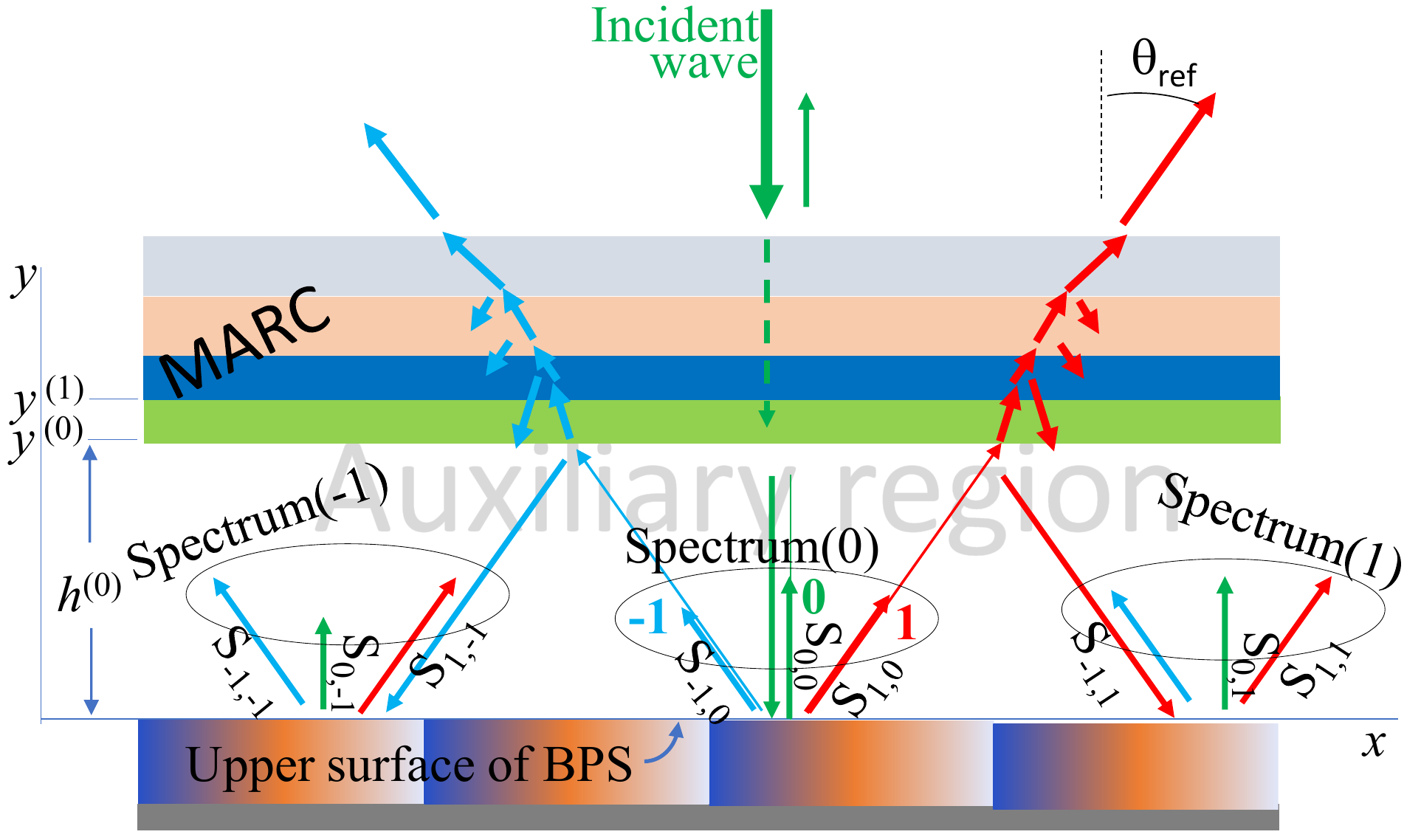}
  \caption{FB wave interactions in an auxiliary region between the MARC and the BPS. In these interactions, multiply-reflected waves $-1$, 0 and 1 (corresponding to the FB indeces of propagating waves in Fig. \ref{FigIntro2}) produce FB discrete Spectrum($-1$), Spectrum(0) and Spectrum(1).  The components of these spectra are related to the terms $S_{p,q}$ of the scattering matrix.} \label{FigForm1}
  \end{center}
\end{figure}

Formulation of the analytical model will be facilitated by employing an auxiliary air region of thickness $h^{(0)}$ between the BPS and the MARC (Fig. \ref{FigForm1}). Eventually, to match the actual configuration, the limit $h^{(0)}\rightarrow 0$ will be taken. The general solutions for the $H$-field in the $m$th MARC layer and in the auxiliary ($m=0$) region are ($y^{(m-1)}<y\leq y^{(m)},0\leq m\leq M$)
 \begin{equation}\label{EqMDARC1}
H^{(m)}_z(x,y)=\sum\limits_{p=-\infty }^{\infty }e^{ik_{xp}x}[A_{p}^{(m)}e^{-ik^{(m)}_{yp}y}+B_{p}^{(m)}e^{ik^{(m)}_{yp}y}], 
\end{equation}
where $y^{(-1)}\equiv 0$. Each term in the sum represents a superposition of a downward wave (the $A_p^{(m)}$ term) and an upward wave (the $B_p^{(m)}$ term) which satisfy the Helmholtz equation in each layer. The unknown coefficients $A_p^{(m)},B_p^{(m)},\rho_p$ may be determined from the boundary conditions that require  continuity of $H_z^{(m)}$ and $E_x^{(m)}=[iZ/(k\epsilon_r^{(m)})]\partial H^{(m)}_z/\partial y$ across each interface, where $Z$ is the impedance of free space. Owing to the orthogonality of the $e^{ik_{xp}x}$ functions over the period $d$, these boundary conditions are satisfied \textit{separately} for each specific FB mode $p$. However, these modes are coupled to each other in the auxiliary region.  There, each of the three propagating modes is multiply reflected between the boundaries at $y=y^{(0)}$ and  $y=0$.  Along the latter boundary at the top of the BPS, each mode is reflected into a spectrum of three propagating modes (Fig. \ref{FigForm1}) with the reflection coefficient of mode $q$ into mode $p$ being related to the scattering matrix coefficient $S_{pq}$ (S-parameters) of the BPS.  Therefore, each “upward” mode with amplitude $B_p^{\text{(0)}}$ is a linear sum of all the “downward” modes with amplitude $A_q^{\text{(0)}}$ which contribute to it \cite[Section S1.5]{SupplementalMat}: 
\begin{equation}\label{EqBPS2}
    B_p^{\text{(0)}}=\sum\limits_{q=-\infty }^{\infty }S_{pq}A_q^{\text{(0)}}.
\end{equation}
	This result is not limited to the propagating rays used for demonstration purposes in Fig. \ref{FigForm1}; it includes evanescent waves as well, which we have found to be essential for the fidelity of the model.   For some surfaces, the $S_{pq}$ can be found in closed form \cite{MarcusPhysRevB2019,MarcusPhysRevB2020,Tretyakov2003,RabinovichIEEETAP2018,RabinovichIEEETAP2020,RabinovichAPL2020}; otherwise, they are readily provided by full-wave commercial programs as by-products of their solution for scattering from a periodic surface \cite{CST}. Hence, we consider them as known for a given BPS, and use them as our starting point for designing the MARC. This further highlights the modularity of our solution: various MARCs can be considered for the same BPS without requiring recalculation of these $S_{pq}$ parameters. 

By truncating the infinite sums in Eqs. \eqref{EqFB1}, \eqref{EqMDARC1} and \eqref{EqBPS2} to sums from $-P$ to $P$ ($P=2$ was found to yield sufficient accuracy for our analytical results), the boundary conditions provide the same number of equations as the number of unknown coefficients. After letting $h^{(0)}\to 0$, this would permit all of the unknowns to be found as a solution to a linear system of simultaneous equations \cite[Section S1.6]{SupplementalMat}, and in particular the coupling efficiencies $\eta_p^\rho$ that we wish to optimize for anoamlous reflection. 

For a given BPS (which generally would not provide the sought-for anomalous reflection), these $\eta^\rho_p$ are clearly related to the physical characteristics of the MARC. In particular, and as mentioned earlier, the modes' cumulative propagation throughout the stratified media, and interaction with the BPS, would determine the result of the modal interference in free space. For simplicity, and because not all substrate materials are commercially available, it will be assumed that the material properties $\epsilon_r^{(m)}$ of each layer, and the number of layers $M$, are given, so that $\eta^\rho_p=\eta^\rho_p(h^{(m)}),1\!\leq\!m\!\leq M$. We utilize this formulation as a basis for solving the inverse problem: determination of the MARC layer thicknesses $h^{(m)}$ which provide the sought-for anomalous reflection values of $\eta^\rho_p$, accomplished via a simple parametric-sweep-based algorithm \cite[Section S2]{SupplementalMat}.       

To demonstrate the proposed scheme and highlight its versatility, we utilize it to realize MARCs for various types of BPSs and anomalous reflection functionalities. It is clear that each additional layer in the MARC provides an additional degree-of-freedom $h^{(m)}$  for attaining the desired anomalous reflection. Since the $\epsilon_r^{(m)}$ are often limited by available inventory, we constrain ourselves to two types of low-loss printed-circuit-board (PCB) substrate materials, arranged in alternate layers: Rogers RO3010 ($\epsilon_r = 10.2$) and RO3003 ($\epsilon_r = 3$) \cite{Rogers}.   All configurations will be investigated for a reflection angle $\theta_\text{ref}=70^\circ$ and a frequency of 20 GHz.   It is apparent, though, that the methodology is entirely applicable to waves in any portion of the electromagnetic spectrum.   

\begin{figure}
  \begin{center}
  \includegraphics[width=3.5in]{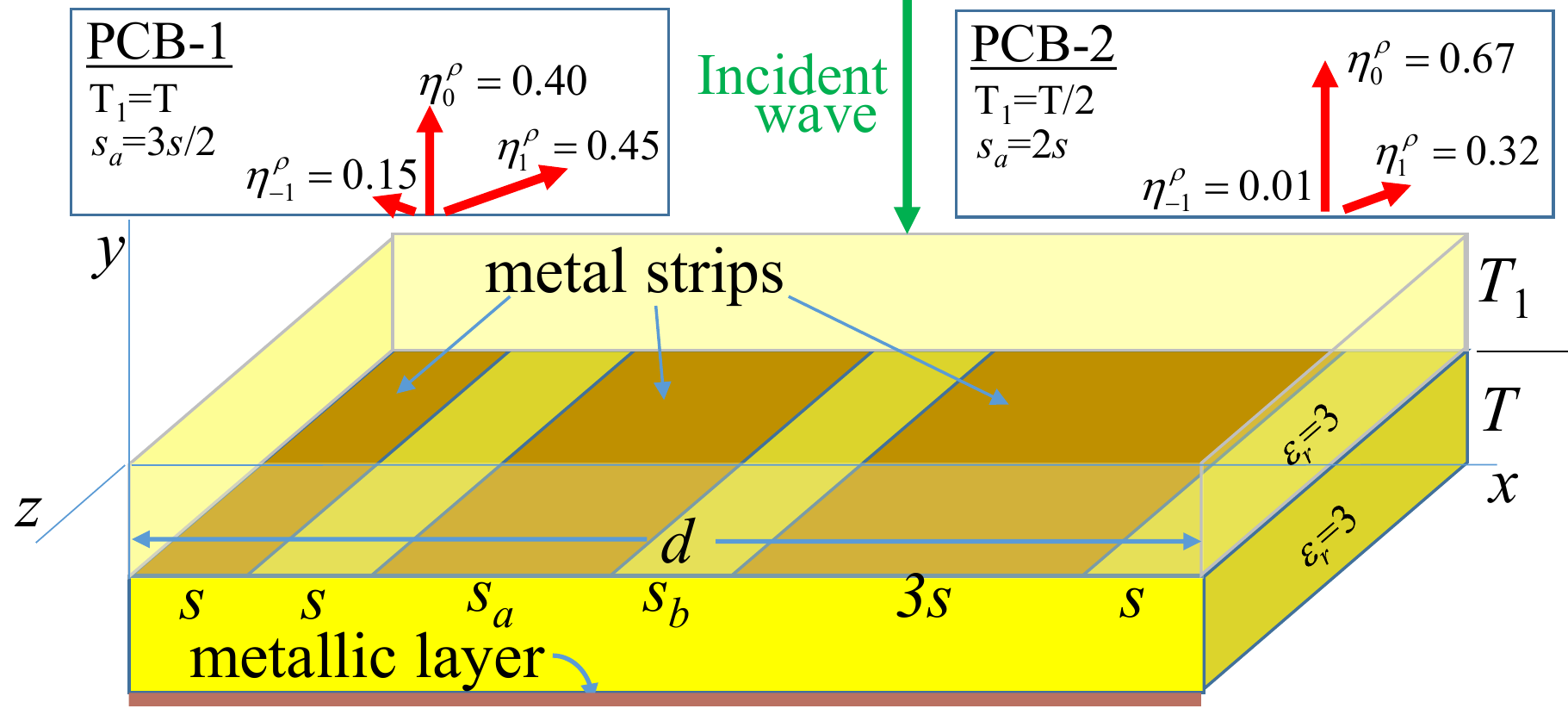}
  \caption{One period of a BPS consisting of three metal strips of widths $s$, $s_a$, $3s$ printed on a grounded dielectric substrate of thickness $T$=60 mil that is covered by a dielectric substrate with the same material of thickness $T_1$. $s=d/9$, $s_b=3s-s_a$. Also shown are the full-wave-calculated mode-coupling results for this BPS alone (i.e. without a MARC), for two sets of values of $T_1$ and $s_a$.} \label{FigPCBBPS}
  \end{center}
\end{figure}

\begin{figure}
  \begin{center}
  \includegraphics[width=3.5in]{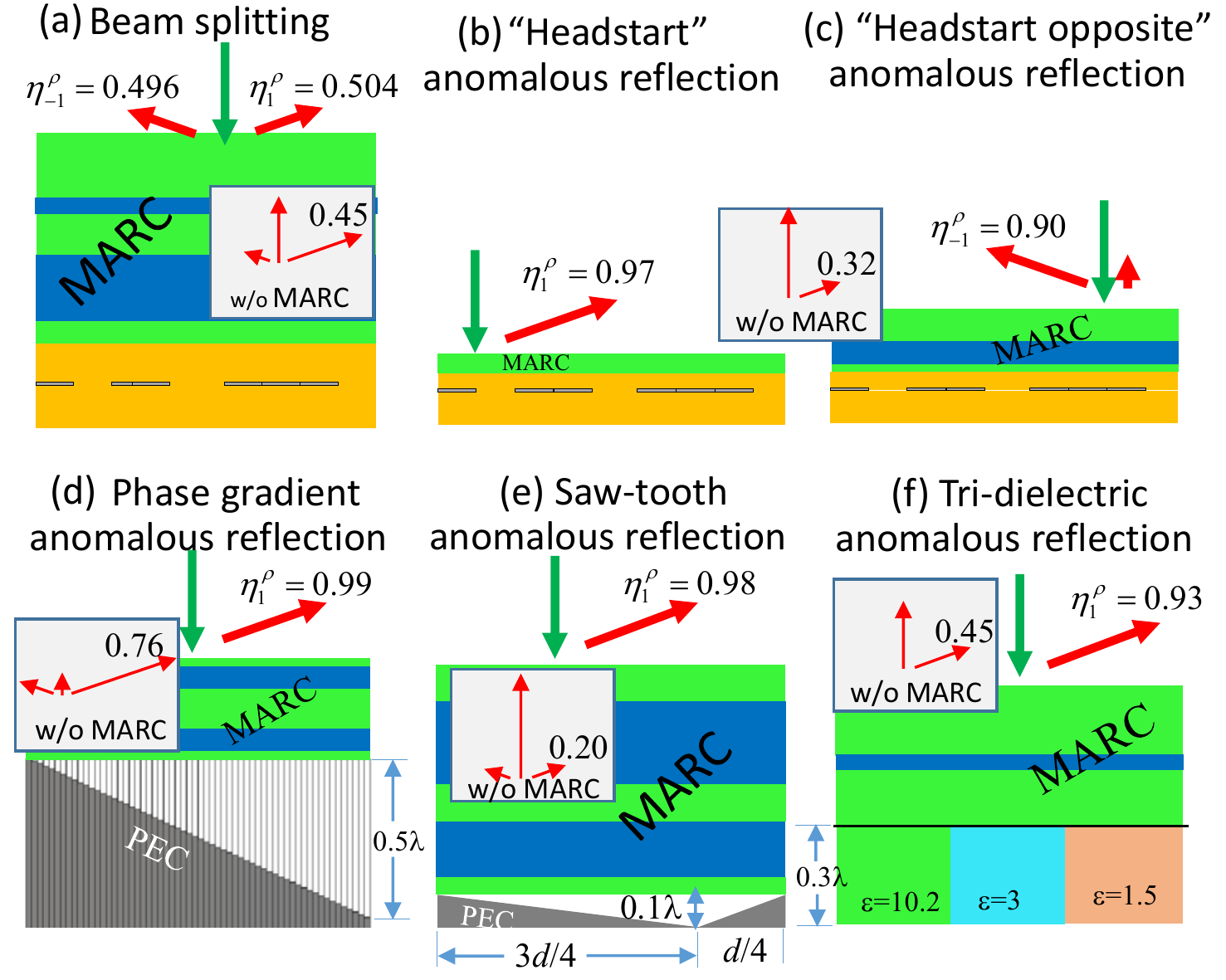}
  \caption{MARC for various purposes and surface-types \cite[Section S4]{SupplementalMat}. For each case, the full-wave-calculated FB spectrum without the MARC is shown as an inset that includes the numerical value of $\eta^{\rho}_1$. The FB spectrum with the MARC is shown above the MARC.  (a) Beam splitting,  PCB-1 BPS of Fig. \ref{FigPCBBPS}; $h^{(m)}\!=\![0.06,0.16,0.11,0.03,0.18]\lambda$. (b) ``Headstart''  anomalous reflection, PCB-2 BPS of Fig. \ref{FigPCBBPS}; $h^{(1)}\!=\!0.058\lambda$. (c) ``Headstart'' opposite-direction anomalous reflection, same BPS as (b); $h^{(m)}\!=\![0.026,0.078,0.107]\lambda$. (d) Anomalous reflection, phase gradient BPS; $h^{(m)}\!=\![0.025,0.075,0.125,0.075,0.025]\lambda$.  (e) Anomalous reflection, bilinear sawtooth-shaped BPS; $h^{(m)}\!=\![0.053,0.179,0.109,0.255,0.094]\lambda$. (f) Anomalous reflection, tri-dielectric BPS; $h^{(m)}\!=\![0.161,0.054,0.211]\lambda$.} \label{FigRes5}
  \end{center}
\end{figure}

We consider first the simple BPS of Fig. \ref{FigPCBBPS} consisting of three metal strips of different widths printed on a grounded dielectric substrate, $\epsilon_r=3$. This configuration is typical of many recent metasurfaces and metagratings at microwave frequencies, thus forming a representative case study for practical applications \cite{MemarianSciRep2017,WongPhysRevX2018,RadiACSPhotonics2018,WongIEEETAP2018,KwonIEEEAWPL2018,WangPhysRevAppl2020}. As part of the BPS, we protect the metal strips  from above by an additional substrate with the same dielectric constant.  This also plays the role of distancing the upper face of the BPS from the thin-metal-induced evanescent waves of higher order, thereby maintaining the validity of the truncation of Eq. \eqref{EqMDARC1} which accounts only for low-order evanescent modes \footnote {Such protective coatings are not always required as seen in Figs. \ref{FigRes5}(d),(e),(f) for non-PCB BPSs.}. Two configurations are considered for the BPS in Fig. \ref{FigPCBBPS}, denoted as PCB-1 and PCB-2 in the figure. These differ in the specific dimensions of both the metallic strips and the protective cover. The corresponding coupling efficiencies $\eta^\rho_p=|S_{p0}|^2$ are shown as well in the inset.        

\begin{figure}
  \begin{center}
  \includegraphics[width=3.5in]{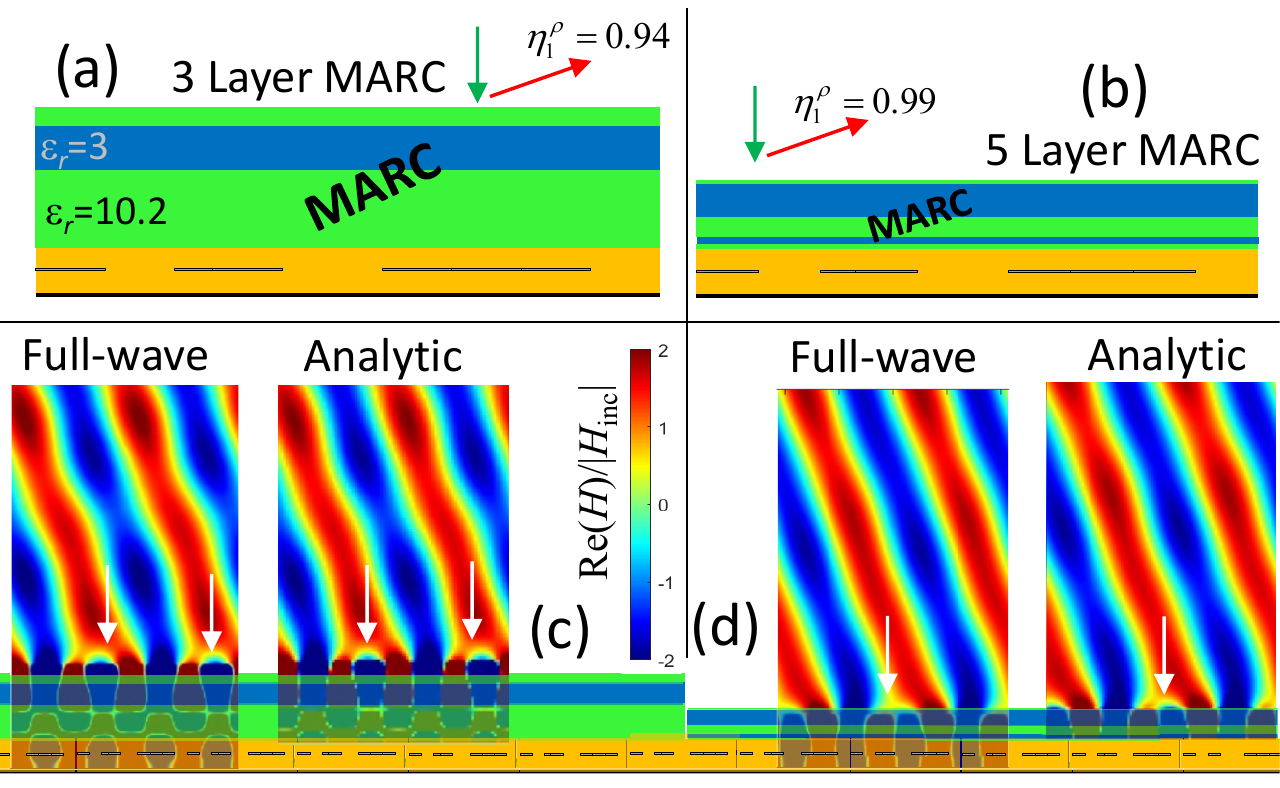}
  \caption{Anomalous reflection solutions for  $M=3$-layer [(a) and (c)] and $M=5$-layer [(b) and (d)] MARC configurations, PCB-1 BPS of Fig. \ref{FigPCBBPS}. (a) and (b): The MARC layers and anomalous reflection results. (c) and (d): $H_z(x,y)$ as computed by the analytical model and by full-wave CST simulations. Green and blue layers in the MARC correspond to Rogers RO3010 ($\epsilon_r=10.2$) and RO3003 ($\epsilon_r=3$), respectively.} \label{FigRes1}
  \end{center}
\end{figure}

 Since the BPS alone does not produce perfect anomalous reflection ($\eta_1^\rho = 0.45$ for PCB-1 and $\eta_1^\rho=0.32$ for PCB-2), it is desired to find a MARC to coat the BPS so that the combined structure achieves $\eta^\rho_1\rightarrow 1,\eta^\rho_{-1}=\eta^\rho_{0}\rightarrow 0$. The PCB-1 configuration for the BPS shown in Fig. \ref{FigPCBBPS} is first considered for $M=3$ [Figs. \ref{FigRes1}(a),(c)] and $M=5$ [Figs. \ref{FigRes1}(b),(d)] MARC layers. Resolving the optimal layer dimensions based on the model and methodology outlined above \cite[Section S2]{SupplementalMat} leads to the enhanced BPS-MARC configurations and resulting anomalous reflection performance shown in Figs. \ref{FigRes1}(a),(b): $h^{(m)}/\lambda=[0.452,0.225,0.104]$, $\eta^\rho_1=0.94$ for $M=3$; $h^{(m)}/\lambda=[0.02,0.04,0.08,0.15,0.01]$, $\eta^\rho_1=0.99$ for $M=5$.  Indeed, as apparent from the plots, utilizing more degrees of freedom (larger $M$) may lead to improved solutions, providing $\eta^\rho_1$ values closer to unity for smaller overall thicknesses. The success of the analytical model may be seen in Figs. \ref{FigRes1}(c) and (d) which compare its $H_z(x,y)$ results to those of CST in each of the media for these two MARCs \footnote{In these figures, the incident wave has been analytically removed from the total fields in the uppermost region. Although the analytically-computed fields are available only above the BPS [$y>0$, see Eq. \eqref{EqMDARC1}], the full-wave-computed fields include the BPS region.}.  For each configuration, the presence of surface waves (white arrows) is apparent. Such surface waves, which are implicitly excited by our BPS-MARC solution, had to be specifically designed into previous solution methods \cite{EpsteinPRL2016,DiazRubioSciAdv2017}. 

%
%
%
The potpourri of BPS-MARC configurations in Fig. \ref{FigRes5} provides insight into the diversity of functionality and BPS-types to which the MARC may be applied \cite[Section S3]{SupplementalMat}.  For example, just as MARCs coated a BPS in Fig. \ref{FigRes1} to produce perfect anomalous reflection in the $\theta_{\text{ref}}$ direction, Fig. \ref{FigRes5}(a) displays a MARC coating the \textit{same} BPS to produce a beam that is "perfectly" split into the $\pm \theta_{\text{ref}}$ directions. 

%

%
The BPS in Figs. \ref{FigRes5}(b) and (c) is the PCB-2 BPS configuration of Fig. \ref{FigPCBBPS} so that for the bare BPS, $\eta^{\rho}_{-1}\approx0$ which is one of the requirements for anomalous reflection. It might be expected that using this BPS would provide a ``headstart'' for anomalous reflection, and that a relatively simple MARC would suffice. Such a simple MARC is indeed shown in Fig. \ref{FigRes5}(b), where only a single layer of relatively minute thickness $0.058\lambda$ provides the anomalous reflection $\eta^{\rho}_{1}=0.97$.  It should be emphasized, though, that this starting point does not prevent us from devising a MARC that would actually reverse the direction of the original beam to the $\eta^{\rho}_{-1}$ direction. This stems from contributions to $\eta^{\rho}_{-1}$ from waves in other directions that only exist in the presence of the MARC (see Fig. \ref{FigForm1}). To demonstrate this, Fig. \ref{FigRes5}(c) displays a MARC, synthesized using the methodology described above, which coats the same PCB-2 BPS, and produces $\eta^{\rho}_{-1}=0.90$; almost enough to be considered \textit{oppositely-directed} anomalous reflection! That is, instead of a "standard" design target $\eta^\rho_1=1$ for which the reflection angle is $70^\circ$, the MARC succeeded in directing the reflection to $-70^\circ$, this despite the fact that without the MARC the BPS produced practically no wave in that direction. 

%
%

%
It should be emphasized that the MARC can provide solutions for \textit{any} type of periodic surface, including complex surfaces composed of many meta-atoms.  Fig. \ref{FigRes5}(d) displays a phase gradient metamaterial consisting of an array of parallel plate waveguides of different lengths \cite{DiazRubioPRB2017,DiazRubioPhysRevApplied2020}. By itself, this structure produces $\eta^{\rho}_1=0.76$.  Coating it with the MARC produces the impressive anomalous reflection $\eta^{\rho}_1=0.99$. Fig. \ref{FigRes5}(e) demonstrates that the BPS on which the MARC is placed need not be smooth. It displays one period of a saw-tooth-shaped conducting surface with a MARC that again produces perfect anomalous reflection ($\eta^{\rho}_1=0.98$). Finally, Fig. \ref{FigRes5}(f) demonstrates that the MARC solution also works well for a dielectric-based BPS, yielding perfect anomalous reflection ($\eta_1^{\rho}=0.97$) for the given "tri-dielectric" composite. For each configuration above, the optimum layer thicknesses and $\eta^{\rho}_p$ values found by the analytical model differed by less than 1\% from the full-wave results, verifying the formulation's fidelity.  

The simulations described above assumed the metal in the BPS is zero-thickness PEC, and the dielectrics are lossless.  If 0.5 oz. copper with realistic conductivity is employed instead \cite{Rogers}, the full-wave-calculated result remains effectively unchanged. While dielectric loss may naturally pose a greater challenge to our scheme, considering its reliance on multiple reflections within the dielectric stack, this can be mostly mitigated by proper inclusion of the expected loss into the analytical model itself \cite[Section S3.1]{SupplementalMat}. 

\begin{figure}
  \begin{center}
  \includegraphics[width=3.5in]{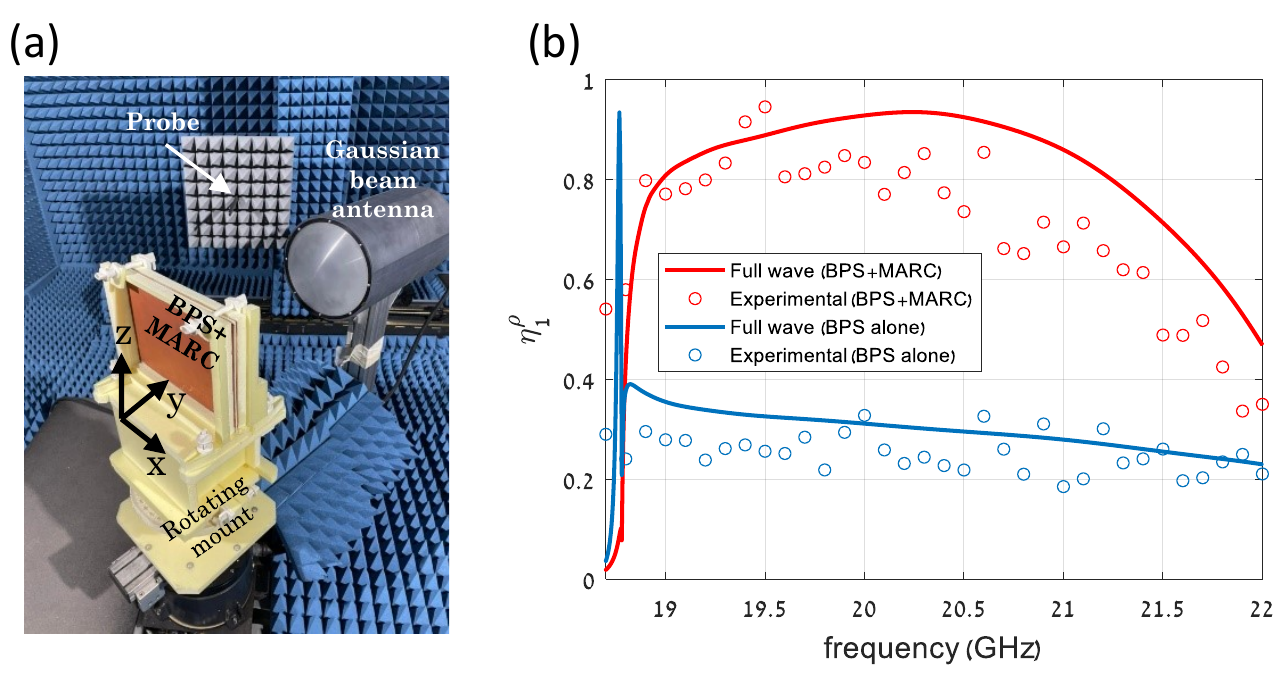}
  \caption{Experimentally measured frequency response for the configuration of Fig. \ref{FigRes5}(b), $h^{(1)}=30$ mil. (a)  Experimental setup. (b) Simulated (solid) and measured (circles) $\eta^{\rho}_1$ with (red) and without (blue) the MARC.  The lowest frequency in the plot is near the cutoff frequency for the plotted mode.} \label{FigRes6}
  \end{center}
\end{figure}

Finally, to provide further support for the practical viability of the proposed concept, we have fabricated the BPS-MARC configuration of Fig. \ref{FigRes5}(c), and characterized it experimentally in a cylindrical near-field measurement setup (Fig. \ref{FigRes6}(a)). Comparing in Fig. \ref{FigRes6}(b) the full-wave simulated (solid) and experimentally measured (circles) $\eta_1^\rho$ as a function of frequency for the bare BPS (blue) and the MARC-enhanced device (red) clearly shows the substantial boost in anomalous reflection efficiency facilitated by the MARC superstate. The measured and simulated results agree very well, with slight discrepancies mainly associated with material parameter tolerances. Overall, the MARC performs very well across the band  \cite[Section S4]{SupplementalMat}, increasing the measured (simulated) fraction of incident power coupled to the anomalous mode from $\eta_1^\rho=0.33$ ($\eta_1^\rho=0.31$) to $\eta_1^\rho=0.83$ ($\eta_1^\rho=0.93$) at the design frequency.

%

It is clear, then, that multimodal anti-reflective coatings (MARCs) can be designed for arbitrary types of periodic surfaces (BPS) to achieve perfect or near-perfect anomalous reflection effects. Despite the simplicity of the concept, we have shown theoretically and experimentally that such a rudimentary stack of homogeneous layers can interfere strongly enough with the complex diffraction pattern of the BPS to significantly change its spectral behavior. The proposed modular solution, circumventing the need for high-resolution structural engineering, merely requires utilization of multiple reflections in slabs with different wave velocities. Thus, it is expected to allow similar "corrections" to beam deflecting devices in other physical fields dominated by wave phenomena, e.g. acoustics or seismics, and provide an appealing route to enhance applications in frequency regimes where subwavelength fabrication is challenging (optics, X-rays).



%

\end{document}